\documentstyle[preprint,tighten,aps,epsf,floats,rotate]{revtex}

\begin{document}
\preprint{\normalsize DOE/ER/40561-84-INT00}
\title
{Three-body recombination in Bose gases with large scattering length}

\author
{ P.F. Bedaque$^{a}$,
  Eric Braaten$^{b}$,
  and H.-W. Hammer$^{b}$
}

\address
{ $^a$~Institute for Nuclear Theory,
 University of Washington, Seattle, WA 98195, USA
~\\$^b$ Department of Physics,
 The Ohio State University, Columbus, OH 43210, USA}

\maketitle

\begin{abstract}
An effective field theory for the 
three-body system with large scattering length
is applied to three-body recombination
to a weakly-bound $s$-wave state in a Bose gas.
Our model independent analysis demonstrates 
that the three-body recombination constant $\alpha$ 
is not universal, but can take any value 
between zero and $67.9 \,\hbar a^4 /m$, where $a$ is the scattering 
length. Other low-energy three-body observables can be
predicted in terms of $a$ and $\alpha$.
Near a Feshbach resonance, $\alpha$ should oscillate between
those limits as the magnetic field $B$ approaches the point where 
$a\to\infty$. In any interval of $B$ over which $a$ 
increases by a factor of 22.7, $\alpha$ should have a zero.
\end{abstract}

\setcounter{page}{1}
\vskip 0.5cm

The successful achievement of Bose-Einstein condensation has 
triggered a large interest in interacting Bose gases.
One of the main factors limiting the achievable density in these 
experiments is the loss of atoms through 3-body 
recombination. Such losses occur when three atoms scatter to form a 
molecular bound state (called a ``dimer'' for brevity) and 
a third atom.
The kinetic energy of the final state particles allows them  
to escape from the trapping potential. This 3-body 
recombination process is interesting in its own right as
it provides a unique window on 3-body dynamics.

The number of recombination events per unit time and volume
can be parametrized as $\nu_{rec}=\alpha n^3$, where $\alpha$
is the recombination constant and $n$ the density of the gas.
The calculation of $\alpha$ in general is a complicated problem,
because it is sensitive to the detailed behavior of the interaction
potential \cite{alphath}. The simplest case is
3-body recombination to a weakly-bound $s$-wave state.
For atoms of scattering length $a$ and mass $m$,
the binding energy of the dimer is $B_d=\hbar^2 /(ma^2)$
and the size of the bound state is comparable to 
$a$. Assuming that $a$ is the only important length scale,
dimensional analysis implies $\alpha={\cal C} \hbar a^4/m$, 
where ${\cal C}$ is dimensionless.
The problem of 3-body recombination to a weakly-bound $s$-wave state
has been studied previously \cite{Fed96,NiM99,EGB99,molcond}.
Fedichev {\it et al.} \cite{Fed96} found that the coefficient ${\cal C}$
has the universal value ${\cal C}=3.9$. Nielsen and Macek \cite{NiM99}
and Esry {\it et al.} \cite{EGB99} found that ${\cal C}$ could take any
value between $0$ and ${\cal C}_{\rm max}\approx 65$.

An ideal means to study the dependence of $\alpha$ on the 
scattering length $a$ is to use Feshbach resonances 
\cite{Festh}, which occur when the energy of a 
spin-singlet molecular bound state is tuned to the energy 
threshold for two separated atoms by 
applying an external magnetic field $B$. Such resonances have, {\it e.g.},
been observed for $^{23}$Na and $^{85}$Rb atoms \cite{Fesex,Ste99}. When the
magnetic field is varied in the vicinity of the resonance, the 
scattering length varies according to \cite{Festh}
\begin{equation}
\label{feshbres}
a(B)=a_0 \left(1+\frac{\Delta_0}{B_0-B} \right)\;,
\end{equation}
where $a_0$ is the off-resonant scattering length and $\Delta_0$ and 
$B_0$ characterize the width and position of the resonance,
respectively. On the side of the resonance where $a$ increases
towards $+\infty$, the spin-singlet molecule becomes a weakly-bound
$s$-wave state.

In this Letter, we use effective field theory methods to calculate
the recombination constant $\alpha$ for a weakly-bound $s$-wave state.
We find that the naive scaling 
relation $\alpha={\cal C} \hbar a^4/m$ is modified by renormalization 
effects, so the coefficient ${\cal C}$ is not universal, and that its 
maximum value is ${\cal C}_{\rm max}=67.9$.
Our results confirm those of Refs. \cite{NiM99,EGB99},
demonstrate that they are model-independent, and provide a new
tool for precise studies of the recombination rate. 
On the side of a Feshbach resonance where $a \to \infty$, 
the $B$-dependence of $\alpha$ can be predicted in terms of
one adjustable parameter.
It has a remarkable behavior, with ${\cal C}$ oscillating
between zero and 67.9 more and more rapidly as $B$ approaches the resonance.

Effective field theory (EFT) is a powerful method for describing
systems composed of particles with wave number $k$ much smaller
than the inverse of the characteristic range $R$ of their interaction.
(For a van der Waals potential $-C_6/r^6$, $R=(2mC_6 /\hbar^2)^{1/4}$).
EFT focuses on the aspects of the problem that are universal, independent
of the details of short-distance interactions,
by modelling the interactions as pointlike.
The separation of scales $k\ll 1/R$ allows a 
systematic expansion in powers of the small parameter $k R$ \cite{gospel}.
Generically, the scattering length $a$
is comparable to $R$, and the expansion is effectively
in powers of $ka$. The pointlike interactions of the EFT generate 
ultraviolet divergences, but they can be absorbed into the 
renormalized coupling constants of the effective Lagrangian.
All information about the influence of short-distance physics on low-energy 
observables is captured by these constants. At any given order in $kR$,
only a finite number of coupling constants enter and this gives the EFT its
predictive power. The domain of validity of EFT is $kR \ll 1$,
even in the case of large scattering length $a \gg R$. Thus it 
should accurately describe weakly-bound states with size of order $a$.
However, the dependence on $ka$ is nonperturbative for $k \sim 1/a$,
and it is necessary to reorganize the perturbative expansion into
a new expansion in $kR$ by resumming higher order terms to all orders 
in $ka$. There has been significant progress recently in carrying out this  
resummation for the 3-body system. At leading order in $kR$,
a single 3-body parameter is necessary and sufficient to carry out the
renormalization \cite{BHK99}. The scattering length and this 3-body
parameter are sufficient to describe all low-energy 3-body observables
up to errors of order $R/a$.
In nuclear physics, this result has allowed a successful description
of low-energy neutron-deuteron scattering and the binding
energy of the triton. The variation of the 3-body parameter
provides a natural explanation for the Phillips line \cite{BHK00}.

We will apply this EFT for 
systems with large scattering length to the 3-body recombination problem. 
For simplicity, we now set $\hbar=1$. We start by writing down a 
general local Lagrangian for a non-relativistic boson field $\psi$:
\begin{equation}
\label{lag}
{\cal L}  =  \psi^\dagger
  \bigg(i\frac{\partial}{\partial t}+\frac{\vec{\nabla}^{2}}{2m}\bigg)\psi
 - \frac{C_0}{2} (\psi^\dagger \psi)^2
 - \frac{D_0}{6} (\psi^\dagger\psi)^3 + \ldots .\nonumber
\end{equation}
The dots denote terms with more derivatives and/or fields;
those with more fields will not contribute to the 3-body amplitude,
while those with more derivatives are suppressed at low
momentum. In order to set up integral equations for 3-body
amplitudes, it is convenient to rewrite ${\cal L}$ by introducing 
a dummy field $d$ with the quantum numbers of two bosons,
\begin{eqnarray}
\label{lagt}
{\cal L}  &=&  \psi^\dagger
\bigg(i \frac{\partial}{\partial t} +\frac{\vec{\nabla}^{2}}{2m}\bigg)\psi
  + d^\dagger d
  -\frac{g}{\sqrt{2}} (d^\dagger \psi\psi +\mbox{h.c.})\nonumber\\
      &+&h\, d^\dagger d \psi^\dagger\psi
 +\ldots\,.
\end{eqnarray}
\noindent
The original Lagrangian (\ref{lag}) is easily recovered by a
Gaussian path integration over the $d$ field, which implies
$d=(g/\sqrt{2})\psi^2/(1+h\psi^\dagger \psi)$, $C_0 = g^2$, and 
$D_0= -3h g^2$. The atom propagator has the usual non-relativistic
form $i/(\omega-p^2/2m)$. The bare propagator for $d$ is simply $i$,
but the exact propagator,
including atom loops to all orders, is \cite{BHK99}
\begin{equation}
\label{Dprop}
i S_d (\omega, \vec{p}) =  \frac {-i 4\pi/(m g^2)}{-1/a
       +\sqrt{-m \omega +\vec{p}^{\,2}/4-i\epsilon}}\,,
\end{equation}
\noindent
where $a$ is the scattering length, which is related to the bare parameter
$g$ and the ultraviolet (UV) cutoff $\Lambda$ by 
\begin{equation}
\label{infrared}
a=\frac{m g^2}{4 \pi} \left(1+\frac{m g^2 \Lambda}{2 \pi^2}
\right)^{-1}\,.
\end{equation}
The propagator (\ref{Dprop}) has a pole at 
$\omega=-1/(ma^2)+\vec{p}^{\,2}/(4m)$ corresponding to a 
weakly-bound state.
Attaching four atom lines to this propagator gives the exact two-atom
scattering amplitude. Taking the incoming atoms to have momenta
$\pm \vec{p}$, the amplitude is $(-1/a-ip)^{-1}$, confirming the 
identification of $a$ as the scattering length.

\begin{figure}[t]
\begin{center}
\epsfxsize=12.cm
\centerline{\epsffile{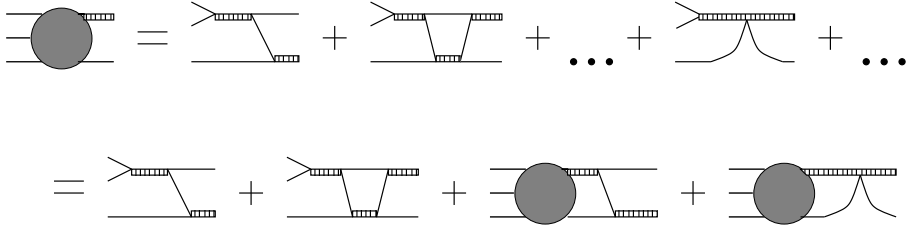}}
\end{center}
\caption{First row: diagrams contributing to 3-body recombination. Second 
row: integral equation summing them to all orders. Atom propagator and
exact dimer propagator are indicated by thin (thick) lines, respectively.}
\label{fig1}
\end{figure}

We now consider the 3-body recombination process.
We take the momenta of the incoming atoms to be small compared to the 
momenta of the final particles, which have magnitude $p_f$.
Using Fermi's golden rule, the recombination coefficient can be 
written 
\begin{equation}
\label{eqalpha}
\alpha= \frac{m a p_f^2}{6\sqrt{3}\pi} \left|T(p_f)\right|^2 \,,
\end{equation}
where $T(p)$ is the amplitude for the transition between three atoms at rest
and a final state consisting of an atom and a dimer in an $s$-wave state
with momentum $p$ in their center-of-momentum frame.
In Eq. (\ref{eqalpha}), this amplitude is evaluated on shell at the 
value $p_f=2/(\sqrt{3}a)$ prescribed by energy
conservation. However, $T(p)$ is also defined at off-shell values of $p$.
The first few diagrams contributing to $T$ are illustrated in Fig. \ref{fig1}.
All loop diagrams are of the same order as the tree diagrams,
and they therefore have to be summed to all orders.
This is conveniently accomplished by solving the integral equation
represented by the second equality in Fig. \ref{fig1}. Corrections to
this equation are of order $R/a$. The integral equation is
\begin{eqnarray}
\label{inteq}
& &T(p)=\frac{96 \pi^{3/2} \sqrt{a}}{m}\left(\frac{1}{p^2}
      +\frac{h}{2 m g^2}\right)+ \frac{2}{\pi}\int_0^\Lambda dq\\
& &      \times\frac{q^2\, T(q)}{-1/a+\sqrt{3}q/2-i\epsilon} 
    \left[\frac{1}{pq}\ln\left|\frac{q^2+p q+p^2}{q^2-q p+p^2}\right|
    +\frac{h}{m g^2}\right]\,,\nonumber
\end{eqnarray}
where we have inserted an UV cutoff $\Lambda$ on the integral over $q$.
If we were allowed to set $h=0$ and take $\Lambda\to\infty$, a rescaling of 
the variables in Eq. (\ref{inteq}) would lead to $T(p)=
K(ap) a^{3/2}/(mp)$, with $K(x)$ a dimensionless function.
Evaluating this in Eq. (\ref{eqalpha}), the
scaling relation $\alpha={\cal C} a^4/m$ would follow immediately.
However, the integral equation (\ref{inteq}) has the same properties as 
the one describing atom-dimer scattering \cite{BHK99} and the 
limit $\Lambda\to\infty$ can {\it not} be taken. 
The individual diagrams are finite as $\Lambda \to \infty$, but 
their sum is sensitive to the cutoff. In an EFT,
the dependence on the UV cutoff is cancelled by
local counterterms. In Ref. \cite{BHK99}, it
was shown that the dependence of the low-energy observables 
on the cutoff $\Lambda$ could be
precisely compensated by varying $h$ appropriately. Writing
$h=2mg^2 H(\Lambda)/\Lambda^2$, it was found that $H(\Lambda)$ could
be well approximated by
\begin{equation}
\label{h}
H(\Lambda)\approx -\tan\left[s_0 \ln(\Lambda/\Lambda_*) -\pi/4 \right]\,,
\end{equation}
where $s_0\approx1.0064$ is determined by the asymptotic behavior
of the integral equation. This expression defines 
a parameter $\Lambda_*$ \cite{BHK99} that characterizes the
effect of the 3-body force on physical observables.
A remarkable feature of this expression is its periodicity in $\ln \Lambda$.
As $\Lambda$ is increased, $H(\Lambda)$
decreases to $-\infty$, changes discontinously to $+\infty$,
and continues decreasing. 

%\vspace{-1.0cm}

\begin{figure}[htb]
\begin{center}
\epsfxsize=10.cm
\centerline{\rotate[r]{\ \ \ \epsffile{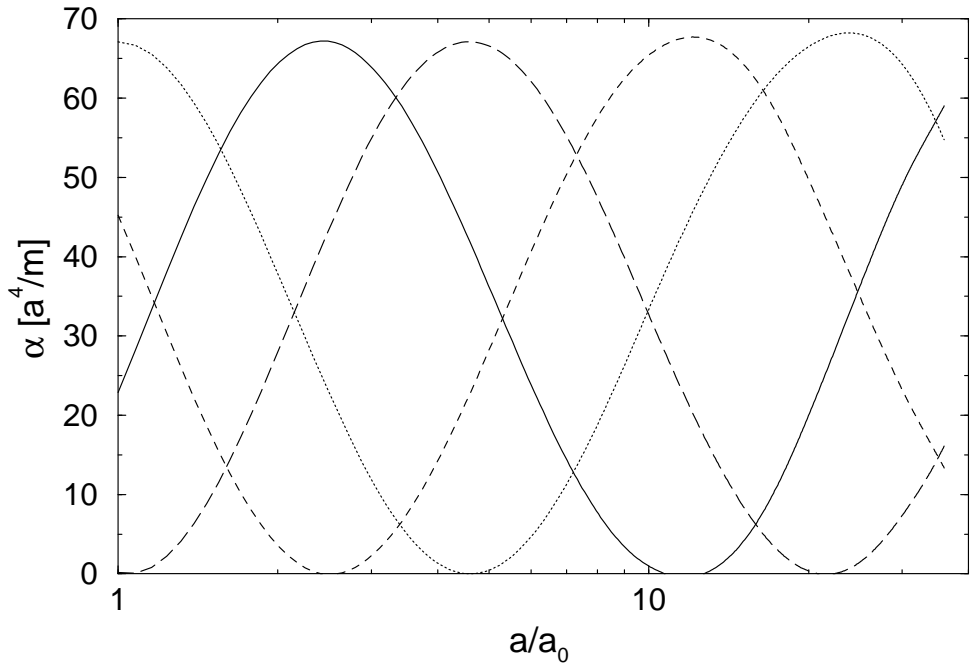}}}
\end{center}
\caption{$\alpha$ as a function of $a$ for different values of the 3-body
parameter: $a_0 \Lambda_*=1.78$ (full line), 4.15 (dotted line), 
7.26 (dashed line), and 19.77 (long-dashed line).}
\label{fig2}
\end{figure}

The scaling violations in $T(p)$ from the
renormalization of the 3-body force can be expressed as a dependence
on $a\Lambda_*$. The simple scaling relation for $\alpha$ is therefore 
replaced by $\alpha = {\cal C} (\Lambda_* a)\, a^4/m$.
Thus the value of ${\cal C}$ is not universal.
In Fig. \ref{fig2}, we show $\alpha$ as a function of $a$
for $\Lambda_* a_0 =1.78,\,4.15,\,7.26$ and 19.77, where $a_0$
is an arbitrary but fixed length scale. 
Interestingly, $\alpha$ appears to oscillate as a function of $\ln a$
between zero and a maximum value ${\cal C}_{\rm max}$. 
We find that the curves can be very well fit by the expression
\begin{equation}
\label{alphaconst}
\alpha\approx\frac{\hbar a^4}{m} {\cal C}_{\rm max} 
\cos^2 \left[s_0\ln(a\Lambda_*)+\delta\right]\,,
\end{equation}
with ${\cal C}_{\rm max}=67.9\pm 0.7$ and $\delta=1.74\pm 0.02$. 

We now compare our result to those of Refs. \cite{Fed96,NiM99,EGB99}.
Using an approximate solution to the 3-body wavefunction
in hyperspherical coordinates, Fedichev {\it et al.} \cite{Fed96} 
found that the coefficient ${\cal C}$ has
the universal value ${\cal C}=3.9$, independent of the interaction 
potential. We find that the value of ${\cal C}$ is
not universal, but can vary from zero to about 67.9, 
depending on the details of 3-body
interactions at short distances. We suggest that the specific value in
Ref. \cite{Fed96} must correspond to an implicit assumption about
the short-distance behavior of the 3-body wavefunction.
Our results are consistent with those of Refs. \cite{NiM99,EGB99},
which were obtained using the hyperspherical adiabatic approximation.
Nielsen and Macek \cite{NiM99} obtained ${\cal C}_{\rm max}\approx
68.4$ by applying the hidden crossing theory. Esry {\it et al.}
\cite{EGB99} used coupled channel calculations to obtain $\alpha$ 
numerically for over 120 different 2-body potentials. For $a\gg R$,
their empirical result has the form of Eq. (\ref{alphaconst})
with $s_0 =1$ and ${\cal C}_{\rm max}= 60 \pm 13$. Refs.
\cite{NiM99,EGB99} show that the zeroes in $\alpha$ arise from
interference effects involving 2-body and 3-body hyperspherical
adiabatic potentials.
The origin of these interference effects is less obvious in our
EFT approach. However, our approach has several other advantages.
First, it is completely model independent. Second, it is a controlled
approximation, with corrections from finite range effects
suppressed by powers of $R/a$. Third, it has predictive power in that
all other low-energy 3-body observables can be determined in terms of 
$a$ and $\Lambda_*$. For example,
the atom-dimer scattering length $a_d$ can be fit by \cite{BHK99}
\begin{equation}
a_d \approx a\left(1.4-1.8\tan[s_0 \ln(a\Lambda_*)+3.2]\right)\,. 
\end{equation}

We now apply the EFT to Feshbach resonances, where
the value of $a$ is varied by changing the external magnetic field 
(cf. Eq. (\ref{feshbres})). Our formalism is only applicable close
to the Feshbach resonance on the side where $a>0$,
because only in that region is there a weakly-bound molecule
with $B_d \sim 1/(ma^2)$. Away from the resonance, or close to the 
resonance but on the side where $a<0$, 3-body recombination must involve 
more deeply-bound molecules with binding energies of order $1/(mR^2)$. 
To predict $\alpha$ as a function of the magnetic field
$B$, we must specify how the parameter $\Lambda_*$ in Eq. (\ref{h})
varies as a function of $B$. A Feshbach resonance is characterized
by nonanalytic dependence of the scattering length $a$ on $B$.
In a field theory, nonanalytic dependence on external parameters
arises from long-distance fluctuations \cite{Wil83}. The explicit 
ultraviolet cutoff $\Lambda$ in our EFT excludes long-distance
effects, which could introduce nonanalytic dependence on $B$,
from the coefficients in ${\cal L}$. Thus the bare parameters
$C_0=g^2$ and $D_0=-3hg^2$ in Eq. (\ref{lag}) should 
be smooth functions of $B$ for a fixed value of $\Lambda$.
The resonant behavior of the scattering length near $B=B_0$ in
Eq. (\ref{feshbres}) can be reproduced by approximating $g^2$ by a
linear function of $B$ in the resonance region. The 
parameter $h$ should also be a smooth function of $B$,
but we can take $h$ to be approximately constant over 
the narrow resonance region.
This assumption implies via Eq. (\ref{h}) that 
$\Lambda_*$ should be constant while $a(B)$ varies as in Eq. 
(\ref{feshbres}).

The behavior of the recombination coefficient near the Feshbach
resonance can therefore be read off from Fig. \ref{fig2}. If $\alpha$
is measured at one value of $B$ for which $a(B) \gg |a_0|$, it
determines $\alpha$ as a function of $B$ up to a two-fold
ambiguity corresponding to whether the slope of ${\cal C}$ is
positive or negative at that value of $B$. 
As $B$ approaches $B_0$, ${\cal C}$ should oscillate 
between 0 and ${\cal C}_{\rm max}$ in the manner shown in Fig. \ref{fig2}. 
The successive zeros correspond to values of $a(B)$ that differ roughly 
by multiplicative factors of $\exp(\pi/s_0)\approx 22.7$. Thus EFT makes the
remarkable prediction that there are values of the magnetic field close to 
a Feshbach resonance where the contribution to $\alpha$ from a 
weakly-bound state vanishes. 

%\vspace{-0.8cm}

\begin{figure}[htb]
\begin{center}
\epsfxsize=10.cm
\centerline{\ \ \ \ \ \ \ \rotate[r]{\epsffile{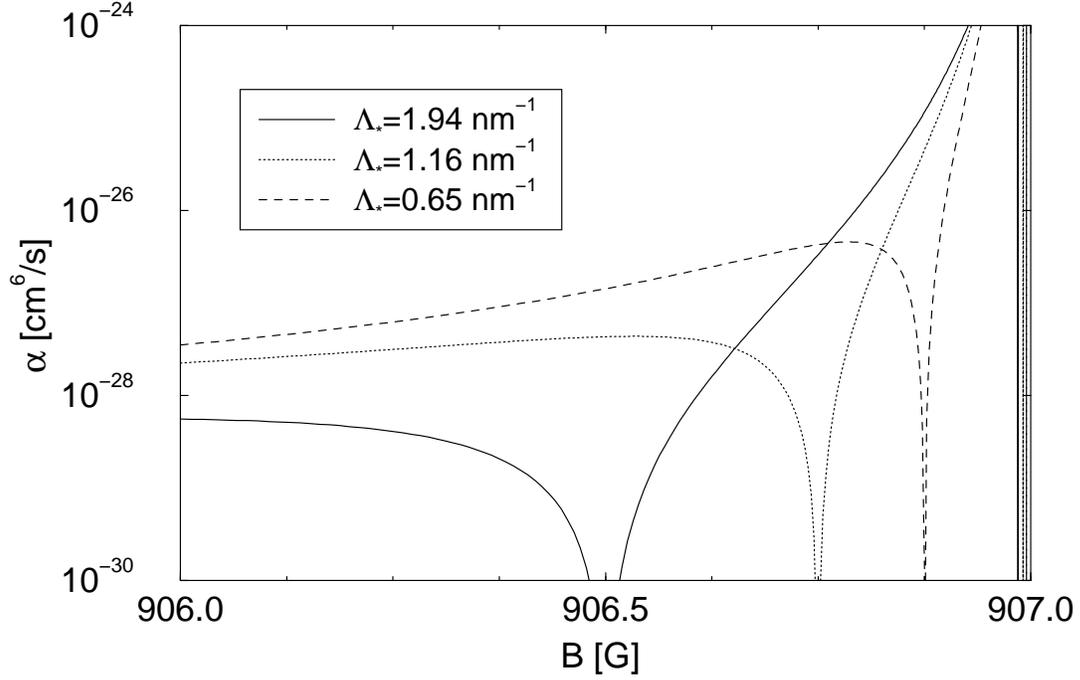}}}
\end{center}
\caption{Contribution to $\alpha$ from recombination into the
weakly-bound state as a function of $B$ for the 907 G Feshbach resonance
in $^{23}$Na.}
\label{fig3}
\end{figure}

The loss rate of $^{23}$Na atoms from a Bose-Einstein condensate
near a Feshbach resonance has been studied by Stenger {\it et al.}
\cite{Ste99}. Our theory applies to the low-field side of the resonance at
$B=907$ G. 
Taking into account the 3 atoms lost per recombination
event and a Bose factor of $1/3!$, the loss rate from the condensate
is $\dot{N}=-\alpha N \langle n^2 \rangle /2$. The loss rates
measured in Ref. \cite{Ste99} correspond to a coefficient 
${\cal C}\approx 300$ both off and near the resonance.
This value is a factor of 4 larger than our maximum value. 
If ${\cal C} > {\cal C}_{\rm max}$, the 3-body recombination rate must be
dominated not by the weakly-bound Feshbach resonance but instead
by molecules with much larger binding energies $\sim 1/(mR^2)$.
Alternatively, the large loss rate in Ref. \cite{Ste99}
could be due to collective effects associated with the Bose-Einstein
condensate, such as a 2-body recombination process involving atoms 
and dimers in a molecular condensate \cite{molcond}. 
In Fig. \ref{fig3}, we show the contribution to $\alpha$ from
the weakly-bound state as a function of the magnetic field
$B$ for $\Lambda_*=1.94 \mbox{ nm}^{-1}$, $1.16 \mbox{ nm}^{-1}$, and 
$0.65 \mbox{ nm}^{-1}$. If this contribution to $\alpha$ could be 
isolated, the first zero may be wide enough to be
observed by experiment. The higher zeroes, however, are increasingly 
narrow and very close to the resonance.

We have applied an EFT for atoms with large scattering length to the 
problem of 3-body recombination into a weakly-bound $s$-wave state.
We find that the coefficient ${\cal C}$ in the scaling relation
$\alpha={\cal C}\hbar a^4/m$ is not universal, but must be in the 
range $0\leq {\cal C}\leq 67.9$. If the 3-body recombination rate 
is measured to be larger than the maximum value, there must be a 
large contribution from molecules that are more deeply bound.
Other low-energy 3-body observables, such as the atom-dimer
scattering length, can be predicted in terms of $a$ and ${\cal C}$. 
Near a Feshbach resonance as $a\to\infty$, we find that
${\cal C}$ should oscillate between zero and 67.9.
In any interval of $B$ over which $a$ increases by a factor of 
$\exp(\pi/s_0)\approx 22.7$,
$\alpha$ should have a zero. Assuming that it is dominated by 
recombination into the weakly-bound state,
the 3-body loss rate should have a minimum at that value of $B$.
If a Bose-Einstein condensate was prepared at such
a value of the magnetic field, one could study its behavior with large 
scattering length and relatively small 3-body losses.

We thank C.H. Greene, G.P. Lepage, J.H. Macek, and U. van Kolck for 
useful discussions.
This research was supported in part by NSF grant PHY-9800964 and
by DOE grants DE-FG02-91-ER4069 and DOE-ER-40561.

%\vspace{-0.4cm}


\begin{thebibliography}{50}
 
%\vspace{-1.cm}

\bibitem{alphath}Yu. Kagan, I.A. Vartan'yants, and G.V. Shlyapnikov,
Sov. Phys. JETP {\bf 54}, 590 (1981); L.P.H. de Goey {\it et al.},
Phys. Rev. B {\bf 34}, 6183 (1986), {\it ibid.} {\bf 38}, 646 (1988);
H.T.C. Stoof {\it et al.}, Phys. Rev. B {\bf 38}, 11221 (1988);
A.J. Moerdijk, H.M.J.M. Boesten, and B.J. Verhaar,
Phys. Rev. A {\bf 53}, 916 (1996).

\bibitem{Fed96}P.O. Fedichev, M.W. Reynolds, and G.V. Shlyapnikov,
Phys. Rev. Lett. {\bf 77}, 2921 (1996).

\bibitem{NiM99}E. Nielsen and J.H. Macek, Phys. Rev. Lett. {\bf 83},
1566 (1999).

\bibitem{EGB99}B.D. Esry, C.H. Greene, and J.P. Burke, Phys. Rev.
Lett. {\bf 83}, 1751 (1999).

\bibitem{molcond}F.A. van Abeelen and B.J. Verhaar, Phys. Rev. Lett.
{\bf 83}, 1550 (1999); V.A. Yurowsky {\it et al.}, Phys. Rev. A {\bf 60},
R765 (1999); E. Timmermans {\it et al.}, {\tt [cond-mat/9805323]}.

\bibitem{Festh}E. Tiesinga, B.J. Verhaar, and H.T.C. Stoof, Phys. Rev. A
{\bf 47}, 4114 (1993); E. Tiesinga {\it et al.}, Phys. Rev. A {\bf 46},
R1167 (1992); A.J. Moerdijk, B.J. Verhaar, and A. Axelsson, Phys. Rev. A
{\bf 51}, 4852 (1995).

\bibitem{Fesex}S. Inouye {\it et al.}, Nature(London) {\bf 392}, 151 (1998);
P. Courteille {\it et al.}, Phys. Rev. Lett. {\bf 81}, 69 (1998); J.L.
Roberts {\it et al.}, Phys. Rev. Lett. {\bf 81}, 5109 (1998).

\bibitem{Ste99}J. Stenger {\it et al.}, Phys. Rev. Lett. {\bf 82}, 2422 (1999).

\bibitem{gospel}
 G.P. Lepage, in {\it From Actions to Answers, TASI'89},
                 ed. T. DeGrand and D. Toussaint,
                 (World Scientific, Singapore, 1990);
 D.B. Kaplan, {\tt [nucl-th/9506035]}.

\bibitem{BHK99}P.F. Bedaque, H.-W. Hammer, and U. van Kolck, Nucl. Phys. A
{\bf 646}, 444 (1999); Phys. Rev. Lett. {\bf 82}, 463 (1999).

\bibitem{BHK00}P.F. Bedaque, H.-W. Hammer, and U. van Kolck, 
Nucl. Phys. A (in print) {\tt [nucl-th/9906032]}; 
Phys. Rev. C {\bf 58}, R641 (1998);
F. Gabbiani, P.F. Bedaque, and H.W. Griesshammer, Nucl. Phys. A
(in print) {\tt [nucl-th/9911034]}. 

\bibitem{Wil83}K.G. Wilson, Rev. Mod. Phys. {\bf 55}, 583 (1983).

\end{thebibliography}
\end{document}